\documentclass[a4paper,pra,twocolumn,showpacs,superscriptaddress,groupedaddress]{revtex4}
\usepackage{ae}
\usepackage[T1]{fontenc}
\usepackage[ansinew]{inputenc}
\usepackage{amsmath}
\usepackage{amssymb}
\usepackage[caption=false]{subfig}
\usepackage{multirow}
\usepackage{epstopdf}
\usepackage{array}
\usepackage[]{graphicx}
\usepackage{wrapfig}
\usepackage{makecell}
\usepackage{color}
\usepackage[colorlinks]{hyperref}
\usepackage{lscape}
\newtheorem{theorem}{Theorem}
\graphicspath{ {./images1/} }

\begin{document}
	\title{Dilution of Entanglement: Unveiling Quantum State Discrimination Advantages}
	\author{Atanu Bhunia}
	\email{atanu.bhunia31@gmail.com}
	\affiliation{Department of Mathematical Sciences, Indian Institute of Science Education and Research Berhampur, Transit Campus, Government ITI, Berhampur 760010, Odisha, India}
	\author{Priyabrata Char}
	\email{mathpriyabrata@gmail.com}
	\affiliation{Department of Applied Mathematics, University of Calcutta, 92, A.P.C. Road, Kolkata- 700009, India}
	 \author{Subrata Bera}
    \email{98subratabera@gmail.com}
    \affiliation{Department of Applied Mathematics, University of Calcutta, 92, A.P.C. Road, Kolkata- 700009, India}
    \author{Indranil Biswas}
    \email{indranilbiswas74@gmail.com}
    \affiliation{Department of Applied Mathematics, University of Calcutta, 92, A.P.C. Road, Kolkata- 700009, India}
\author{Indrani Chattopadhyay}
	\email{icappmath@caluniv.ac.in}
	\affiliation{Department of Applied Mathematics, University of Calcutta, 92, A.P.C. Road, Kolkata- 700009, India}
	\author{Debasis Sarkar}
	\email{dsarkar1x@gmail.com, dsappmath@caluniv.ac.in}
	\affiliation{Department of Applied Mathematics, University of Calcutta, 92, A.P.C. Road, Kolkata- 700009, India}
	   \begin{abstract}
		 The states in the three-qubit GHZ SLOCC class can exhibit diverse entanglement patterns, as they may have no entanglement in any reduced subsystems, or show entanglement across one, two, or all three bipartite cuts. Significant research has explored how such states can be used in entanglement-assisted discrimination tasks. In this paper, we analyze the relationship between probability of error and amount of bipartite and multiparty entanglement, examining how different levels of entanglement impact the accuracy of state discrimination. Also we have shown that the generic class of GHZ state provide some advantages in probabilistic distinguishibility. However, perfect discrimination typically requires maximally entangled states. The use of non-maximally entangled states as a resource for perfect discrimination remains an open problem in this area of research. In this manuscript, we propose a method to explore the perfect distinguishability of orthogonal product states using non-maximally entangled states, utilizing the GHZ SLOCC class structure. Moreover, these findings offer deeper insights into the relationship between entanglement classification and nonlocality, potentially shedding light on how different entanglement structures influence nonlocal behavior in quantum systems.
	\end{abstract} 
	
	\date{\today}
	\maketitle
	
	
	\section{INTRODUCTION}
  	The affinity of quantum entanglement and quantum nonlocality always appraised as a fundamental issue in quantum information theory during former past. Besides that quantum nonlocality has a separate form of manifestation from their own perspective. Infact, when involved systems are quantum, several counterintuitive results occur which are not present in the classical case. For instance, the classical information encoded in the states of a composite quantum system may not be completely extracted under local quantum operations along with classical communications (LOCC) among the spatially separated subsystems. Such a set of states is called nonlocal due to their local discrimination tasks and optimum discrimination of which requires only global operation(s) on the composite system\cite{Bennett1999,bennett1996,popescu2001,xin2008,Walgate2000,Virmani,Ghosh2001,Groisman,Walgate2002,Divincinzo,Horodecki2003,Fan2004,Ghosh2004,Nathanson2005,Watrous2005,Niset2006,Ye2007,Fan2007,Runyo2007,somsubhro2009,Feng2009,Runyo2010,Yu2012,Yang2013,Zhang2014,somsubhro2009(1),somsubhro2010,yu2014,somsubhro2014,somsubhro2016}. For bipartite local discrimination tasks, a state secretly chosen from a set of pre-specified orthogonal quantum states  and our goal is to locally figure out the exact identity of this state. Local quantum state discrimination plays an prominent role in exploring the restrictions of LOCC. In some extent, we have known that entanglement escalates the trouble of state discrimination under LOCC. However, entanglement is not an essential feature for such kind of nonlocality\cite{Bennett1999,Zhang2015,Wang2015,Chen2015,Yang2015,Zhang2016,Xu2016(2),Zhang2016(1),Xu2016(1),Halder2019strong nonlocality,Halder2019peres
  	set,Xzhang2017,Xu2017,Wang2017,Cohen2008,Zhang2019,somsubhro2018,zhang2018,Halder2018,Yuan2020,Rout2019,bhunia2020,bhunia2022,atanu2023,atanu2024,subrata2024,indranil2023}. In 1999, Bennett et al. \cite{Bennett1999} in their seminal paper first constructed an examples of orthonormal product basis in $3\otimes3$ that are not perfectly distinguishable under LOCC. The construction was quite striking as they contain only product states and, hence, introduced the phenomenon of quantum nonlocality without entanglement. Furthermore the entity of incomplete bases which evince the phenomenon of nonlocality without entanglement, familiar as unextendible product bases (UPB) which means a set of mutually orthogonal product states satisfies the condition that no product state lies in the orthogonal complement of the subspace by these states\cite{Divincinzo}. UPB cannot be distinguished perfectly by LOCC, and the projector onto that orthogonal complement is a mixed state which shows the fascinating phenomenon known as bound entanglement. Thus, these states are of considerable interest in quantum information theory.\\
    At the very basis of entanglement theory the paradigm of Local Operations and Classical Communication (LOCC) formulated by Bennett et al \cite{Bennett1996A}. Under this paradigm, a state is shared between different parties who can perform arbitrary local operations (including measurements and operations involving additional local subsystems, so-called ancillas), and in addition can communicate with each other over normal classical channels; however, they are not able to exchange quantum systems. Under such kind of constraints, state transformations become restricted. In particular, any arbitrary state can not be created by following this rule. This provides some categorization between states; a state is separable if it can be created using only local operations and classical communications, and it is called entangled otherwise \cite{Brub2002,Plenio1998}. The fact that entangled states cannot be generated locally makes them a resource \cite{popescu1999}. This resource is used in different tasks in quantum computation, quantum communication and quantum cryptography\cite{Divincenzo1995}.
     
It is well known that entanglement is a very valuable resource allowing remote parties to communicate such as in teleportation \cite{Bennett1993}. The locally indistinguishable quantum states may become distinguishable with a small number of entanglement resources. In 2008, Cohen \cite{Cohen2008} first presented     a method which uses entanglement more efficiently than teleportation to distinguish certain classes of UPBs, and showed that certain classes of UPBs in $m\otimes n (m\leq n)$ can be locally distinguished by using an $\lceil \frac{m}{2}\rceil\otimes \lceil \frac{m}{2}\rceil$ maximally entangled state. Moreover, Cohen \cite{Cohen2008} presented a method to locally distinguish a set of five states on a $3\otimes 3$ bipartite system with only one ebit of entanglement, and it remained an open question whether it is possible to distinguish these states with less than one ebit of entanglement. Later on, entanglement as a resource for local state discrimination in multipartite systems attracted more and more attention \cite{bhunia2020,zhang2018,Rout2019,Cohen2008,atanu2023,atanu2024,subrata2024,indranil2023}.
    
    The states which belongs to three qubit GHZ SLOCC class either have no entanglement in any reduced subsystem as well as can have entanglement in one, two or three bipartite cut. Till now there are many works in the region of entanglement assisted discrimination. But for each case of perfect discrimination, a maximally entanglement is needed. So by using a non-maximally entangled state as a resource for perfect discrimination is still open. In this paper, we propose a method to study the local distinguishability of orthogonal product state using non-maximally entangled state with the help of the structure of GHZ SLOCC class, and analyze the relationship between the probability of error and the quantity of bipartite as well as multipartite entanglement. These results will provide a better understanding of the relationship between entanglement classification and nonlocality.
    
     The remaining portion of this paper is arranged as follows. In Sec.~\ref{A1} necessary definitions and other preliminary concepts are presented. In Sec.~\ref{A2}, we represent the classification of three qubit GHZ class of states. These states, characterized by profound entanglement properties, play a pivotal role in quantum information processing. In Sec.~\ref{A3}, we have succeeded in distinguishing a nonlocal set of orthogonal product states by various class of GHZ states as a resource. Also we are able to illuminate the intricate relationship between entanglement measures and the efficacy of quantum state discrimination. In Sec.~\ref{A4}, we provide an advantage of generic class of states in probabilistic distinguishibility. Finally, the conclusion is drawn in Sec.~\ref{A5} with some open problems for further studies.
	
	\section{Preliminaries}
	\label{A1}
	
Entanglement as a resource raises two central questions: how much is required for a task, and how many types exist. Quantification leads to entanglement monotones \cite{vidal2000} and measures \cite{plenio1997}, while classification becomes essential in multipartite systems and even relevant in bipartite settings.

For two-qubit states $\rho_{AB}$, concurrence \cite{coffman2000} is given by
\begin{equation}
	C(\rho_{AB}) = \max\{0, \lambda_1 - \lambda_2 - \lambda_3 - \lambda_4\},
\end{equation}
where $\lambda_i$ are the square roots of the eigenvalues of $\rho_{AB} (\sigma_y \otimes \sigma_y) \rho_{AB}^* (\sigma_y \otimes \sigma_y)$ in decreasing order. For pure states,
\begin{equation}
	C(\rho_{AB}) = 2 \sqrt{\det \rho_A},
\end{equation}
with $\rho_A = \mathrm{Tr}_B(\rho_{AB})$. The tangle is defined as
\begin{equation}
	\tau = C^2(\rho_{A|BC}) - C^2(\rho_{AB}) - C^2(\rho_{AC}).
\end{equation}

We focus on distinguishing pure orthogonal product states shared among distant parties restricted to LOCC. States are equally probable and pairwise orthogonal, enabling perfect discrimination. However, this requires that orthogonality is preserved across measurement rounds. A quantum measurement on a $d$-dimensional system uses POVM elements $\{\pi_l\}$ satisfying
\begin{equation}
	\sum_l \pi_l = I_{d \times d},
\end{equation}
where $I_{d \times d}$ is the identity matrix. Throughout, we use unnormalized states and operators for notational simplicity.

\textbf{Definition 1} \cite{Halder2018}: A measurement is trivial if all POVM elements are proportional to identity; otherwise, it is nontrivial.

\textbf{Definition 2} \cite{Halder2018}: A measurement is orthogonality-preserving if post-measurement states remain mutually orthogonal.

If no party can begin with a nontrivial, orthogonality-preserving measurement, it is impossible to eliminate any state locally, indicating the nonlocality of the set.

In the next section, we examine GHZ states, their classification, and entanglement structure, which are key to understanding multipartite nonlocality and its implications in quantum information.
\section{GHZ States: Classification and Characteristics}
\label{A2}
 Genuine tripartite entanglement refers to a form of quantum entanglement that involves three or more particles, where the entanglement cannot be reduced to the entanglement between any pair of subsystems. This kind of entanglement goes beyond the traditional bipartite entanglement seen in two-particle systems, as it involves correlations that cannot be fully described by considering only the interactions between subsystems. One of the most notable examples of genuine tripartite entanglement is the Greenberger-Horne-Zeilinger (GHZ) state, given by  $|G H Z\rangle=\frac{1}{\sqrt{2}}(|000\rangle+|111\rangle)$. The GHZ state is a three-qubit generalization of the Bell state, which is central in the study of bipartite entanglement. In contrast to bipartite states, where entanglement properties can be largely captured by focusing on two-qubit subsystems, the GHZ state exhibits correlations that are nonlocal across all three qubits, meaning that the entanglement is truly shared among the entire system, not just between pairs of qubits. This highlights the unique and intricate nature of tripartite entanglement, which plays a critical role in various quantum information protocols, including quantum teleportation, secret sharing, and error correction in quantum computing. The distinctive properties of such states are crucial for advancing our understanding of quantum systems and exploring new ways of harnessing quantum mechanics for practical applications.\\
 
 A state belonging to the GHZ class can be expressed, up to local unitary transformations (LUs), as $|\psi(a,b,c,r)\rangle=\frac{1}{\sqrt{k}} g_x^1 \otimes g_x^2 \otimes g_x^3 P_z(|000\rangle+|111\rangle)$, $P_r=$ $\left(\begin{array}{cc}r & 0 \\ 0 & 1 / r\end{array}\right), r \in \mathbb{C}$ with $|r| \leq 1$ and $\frac{1}{\sqrt{k}}$ is the normalizing factor, $k=\frac{1+|r|^4+\left(r^2+\left(r^*\right)^2\right)\cos a\cos b\cos c}{8|r|^2}$. Now, if we take $g_x^1=\left(\begin{array}{cc}\frac{1}{\sqrt{2}} & \frac{1}{\sqrt{2}}\cos a \\ 0 & \frac{1}{\sqrt{2}}\sin a\end{array}\right)$, $g_x^2=\left(\begin{array}{cc}\frac{1}{\sqrt{2}} & \frac{1}{\sqrt{2}}\cos b \\ 0 & \frac{1}{\sqrt{2}}\sin b\end{array}\right)$, $g_x^3=\left(\begin{array}{cc}\frac{1}{\sqrt{2}} & \frac{1}{\sqrt{2}}\cos c \\ 0 & \frac{1}{\sqrt{2}}\sin c\end{array}\right)$, where $a,b,c\in(0,\frac{\pi}{2}]$. we see that $\left(g_x^1\right)^{+}\left(g_x^1\right)=\left(\begin{array}{cc}\frac{1}{2} & \frac{1}{2} \cos a \\ \frac{1}{2} \cos a & \frac{1}{2}\end{array}\right)=\frac{1}{2} I+\left(\frac{1}{2} \cos a\right) \sigma_x,$ hence $g_1=\frac{1}{2} \cos a \in\left[0, \frac{1}{2}\right)$. Similarly $g_2=\frac{1}{2} \cos b\left[0, \frac{1}{2}\right)$ and $g_2=\frac{1}{2} \cos b\left[0, \frac{1}{2}\right)$. Now the concurrences of the reduced states of $|\psi(a,b,c,r)\rangle$ are
	$$
	\begin{aligned}
		& C_{12}=\frac{2|z|^2 \sin a \sin b \cos c}{1+|z|^4+\left(z^2+\left(z^*\right)^2\right) \cos a \cos b \cos c} \\
		& C_{13}=\frac{2|z|^2 \sin a \cos b \sin c}{1+|z|^4+\left(z^2+\left(z^*\right)^2\right) \cos a \cos b \cos c} \\
		& C_{23}=\frac{2|z|^2 \cos a \sin b \sin c}{1+|z|^4+\left(z^2+\left(z^*\right)^2\right) \cos a \cos b \cos c} \\
		&
	\end{aligned}
	$$
Throughout the Section we establish the classification for GHZ class of non generic pure states. The non generic class is of special interest because the reduced bipartite system of it may or may not have entanglement which depends on the parameters $a, b$ and $c$ as evident from the concurrences. Including all cases they all have non zero tangle. As there exists strong correspondence between the parameters and the entanglement of the reduced system. We now characterize all the sub classes of the non generic class.\\
{\bfseries Case-I} (a=$\pi$/2,b=$\pi$/2,c=$\pi$/2) This is the most simple subclass structure of the non generic GHZ class as there does not exists any kind of bipartite entanglement in the reduced systems. Even from the concurrence relation we can check that $C_{ij}=0$ $ \forall i,j$.\\
{\bfseries Case-II} (a=$\pi$/2,b=$\pi$/2,c$\neq$$\pi$/2) In this non generic GHZ class two concurrences are vanished. So the existence of entanglement of the reduced systems as well as the structure of the class have a notable change from the previous subclass. The concurrences are given $C_{13}=0$, $C_{23}=0$, $C_{12}=\frac{\cos c\sin a\sin b}{4k}$.\\	
{\bfseries Case-III} (a=$\pi$/2,b$\neq$$\pi$/2,c$\neq$$\pi$/2) The complexity of the non generic GHZ class in this case gradually increase. In this case there exists bipartite entanglement of reduced systems in two bipartition. The concurrences of the reduced states are given by,\\
$C_{12}=\frac{\cos c\sin a\sin b}{4k}$, $C_{13}=\frac{\cos c\sin a\sin b}{4k}$, $C_{23}=0$.\\

In the upcoming section, we will explore how the utilization of specific non-generic subclasses of GHZ states can significantly improve the process of quantum state discrimination in multipartite systems. This approach provides a more efficient framework for distinguishing between quantum states, enhancing the accuracy and performance of quantum information protocols within complex multipartite configurations.
\section{Enhancing Quantum State Discrimination in Multipartite Systems via Entanglement}
\label{A3}
	Consider a tripartite quantum system associated with a Hilbert space $\mathcal{H}=\left(\mathbb{C}^3\right)^{\otimes 3}$. In the following, an explicit construction of a small set of pure orthogonal product states $\left\{\left|\psi_i\right\rangle\right\} \in \mathcal{H}=\left(\mathbb{C}^3\right)^{\otimes 3}$ is presented with $i=1, \ldots, 12$. The states are given by
\begin{multline}
	\label{A}
$$
	\;\;\;\;\;\;\;\;\;\;\;\left|\psi_1\right\rangle=|0\rangle|1\rangle|0+1\rangle, \left|\psi_2\right\rangle=|0\rangle|1\rangle|0-1\rangle,\\
	\left|\psi_3\right\rangle=|0\rangle|2\rangle|0+2\rangle, \left|\psi_4\right\rangle=|0\rangle|2\rangle|0-2\rangle,\\
	\left|\psi_5\right\rangle=|1\rangle|0+1\rangle|0\rangle, \left|\psi_6\right\rangle=|1\rangle|0-1\rangle|0\rangle,\\
	\left|\psi_7\right\rangle=|2\rangle|0+2\rangle|0\rangle, \left|\psi_8\right\rangle=|2\rangle|0-2\rangle|0\rangle,\\
	\left|\psi_9\right\rangle=|0+1\rangle|0\rangle|1\rangle, \left|\psi_{10}\right\rangle=|0-1\rangle|0\rangle|1\rangle,\\
	\left|\psi_{11}\right\rangle=|0+2\rangle|0\rangle|2\rangle, \left|\psi_{12}\right\rangle=|0-2\rangle|0\rangle|2\rangle.
$$
\end{multline}
{\bfseries Lemma 1}\cite{Halder2018}  {\it No state from (\ref{A}) can be eliminated by performing orthogonality-preserving measurements.
}\\\\ 
The twisted states \cite{Niset2006} play an important role in local distinguishability of orthogonal product states. Using entanglement as a resource to discriminate a set of locally indistinguishable states, we need sufficient ancillary subsystems \cite{Cohen2008,Rout2019}. For the discrimination task with support of a two qubit Bell state, the dimension of each ancillary subsystem is two and the number of systems holding  ancillary subsystem is also two. To generate a discrimination protocol, sharing Bell state, respective parties chooses appropriate measurements on his/her system together with the ancillary subsystems. Whereas for the discrimination task of a tripartite system with support of a three qubit GHZ state, the dimension of each ancillary subsystem is two but the number of systems holding ancillary subsystem is three. So the choice of measurement for Bell state assisted discrimination task must be a subset of the choice of measurement for GHZ state assisted discrimination task. We summarize the above discussions by writing the following proposition.\\

{\bfseries Lemma 2} {\it In a tripartite system if a set of orthogonal product states is distinguishable by using a two qubit Bell state shared between any two parties, then the set of states can be distinguished by using a three qubit GHZ state. But the converse is not true always.}\\

 For $3\otimes3\otimes3$ system, the set of states (\ref{A}) can be distinguished by sharing one copy of Bell state shared between any two parties \cite{Halder2018}. So by the previous Lemma it must be distinguished by one copy of GHZ state.\\\\
{\bfseries Corollary 1} {\it One copy of three-qubit GHZ is sufficient to distinguish the set of states (\ref{A}).}\\
$Proof:$ First, we assume that a GHZ state shared between three parties Alice, Bob and Charlie be $|\psi\rangle_{abc}$. Then the initial states shared among them is $$\left|\phi\right\rangle_{ABC}\otimes\left|\psi\right\rangle_{abc}$$
where $\left|\phi\right\rangle$ is one of the state from (\ref{A}).\\

Step $1 .$ Alice performs a measurement
$$
\begin{aligned}
	\mathcal{A} & \equiv\left\{M:=\mathbb{P}\left[(|1\rangle,|2\rangle)_{A} ;|1\rangle_{a}\right]+\mathbb{P}\left[|0\rangle_{A} ;|0\rangle_{a}\right]\right.,\\
	\bar{M} &:=\mathbb{I}-M\}
\end{aligned}
$$
Suppose the outcomes corresponding to $M$ clicks. The resulting post-measurement states are therefore,
\begin{multline*}
	$$
	\left|\psi_{1},\psi_{2}\right\rangle\rightarrow|0\rangle_{A}|1\rangle_{B}\left|0 \pm 1\rangle_{C}|000\rangle_{abc}\right.,\\
	\left|\psi_{3},\psi_{4}\right\rangle\rightarrow|0\rangle_{A}|2\rangle_{B}\left|0 \pm 2\rangle_{C}|000\rangle_{abc}\right.,\;\;\;\;\;\;\;\;\;\;\;\;\;\;\;\;\;\;\;\;\;\;\;\;\;\;\;\;\;\;\\
	\left|\psi_{5},\psi_{6}\right\rangle\rightarrow|1\rangle_{A}\left|0 \pm 1\rangle_{B}|0\rangle_{C}|111\rangle_{abc}\right.,\;\;\;\;\;\;\;\;\;\;\;\;\;\;\;\;\;\;\;\;\;\;\;\;\;\;\;\;\;\;\\
	\left|\psi_{7},\psi_{8}\right\rangle\rightarrow|2\rangle_{A}\left|0 \pm 2\rangle_{B}|0\rangle_{C}|111\rangle_{abc}\right.,\;\;\;\;\;\;\;\;\;\;\;\;\;\;\;\;\;\;\;\;\;\;\;\;\;\;\;\;\;\;\\
	\left|\psi_{9},\psi_{10}\right\rangle\rightarrow|0\rangle_{A}\left|0\rangle_{B}|1\rangle_{C}|000\rangle_{abc}\right.\pm|1\rangle_{A}\left|0\rangle_{B}|1\rangle_{C}|111\rangle_{abc}\right.,\\
	\left|\psi_{11},\psi_{12}\right\rangle\rightarrow|0\rangle_{A}\left|0\rangle_{B}|2\rangle_{C}|000\rangle_{abc}\right.\pm|2\rangle_{A}\left|0\rangle_{B}|2\rangle_{C}|111\rangle_{abc}\right.,\\
	$$
\end{multline*}
Step $2 .$ Bob performs a measurement

If the outcomes corresponding to $N_1$ and $N_2$ clicks. The resulting post measurement states are therefore $|0\rangle_{A}|1\rangle_{B}\left|0 \pm 1\rangle_{C}|000\rangle_{abc}\right.$ and $|0\rangle_{A}|2\rangle_{B}\left|0 \pm 2\rangle_{C}|000\rangle_{abc}\right.$ respectively, which can be easily distinguished by Charlie by projecting onto $|0\pm1\rangle_{C}, |1\pm2\rangle_{C}$ . If the outcome corresponding to $\bar{N}$ clicks, then the remaining eight states are isolated.\\
Step $3 .$ Charlie performs a measurement
$$
\begin{aligned}
	\mathcal{C} \equiv\{&Q_1:=\mathbb{P}[|1\rangle_{C} ;|0\rangle_{c}]+\mathbb{P}[|1\rangle_{C} ;|1\rangle_{c}],\\
	&Q_2:=\mathbb{P}[|2\rangle_{C} ;|0\rangle_{c}]+\mathbb{P}[|2\rangle_{C} ;|1\rangle_{c}],\\
	&\bar{Q} :=\mathbb{I}-Q_1-Q_2\}
\end{aligned}
$$		
there are only two states due to each outcomes $Q_1$ and $Q_2$. For each cases  Walgate et.al. \cite{Walgate2000} result follows. If the outcome corresponding to $\bar{Q}$ clicks, then the remaining four states are isolated. Afterthat Alice makes two outcome rank one measurement $|1\rangle_A|1\rangle_a$ and $|2\rangle_A|1\rangle_a$.there are only two states due to each outcomes and which can be easily distinguished by Bob easily. Hence those states are perfectly distinguished.\\

\begin{theorem}
	\label{th1}
Let an ancillary nonmaximally entangled state shared between Alice and Bob, such as $|\chi\rangle=\frac{r}{\sqrt{1+r^4}}(r|00\rangle_{ab}$+$\frac{1}{r}|11\rangle_{ab})$, |r|$\leq$1. Then (\ref{A}) can be locally distinguished with probability of error $p_e=\frac{1-r^4}{1+r^4}\xi_e$, $0<\xi_e<1$.
\end{theorem}

{\it Proof:}  Suppose an ancillary nonmaximally entangled state shared between Alice and Bob, such as $|\chi\rangle=\frac{r}{\sqrt{1+r^4}}(r|00\rangle_{ab}$+$\frac{1}{r}|11\rangle_{ab})$. Then the initial states are transformed into
$$
\left|\psi_{i}^{\prime}\right\rangle=\left|\psi_{i}\right\rangle \otimes\left[\frac{r}{\sqrt{1+r^4}}(r|00\rangle_{ab}+\frac{1}{r}|11\rangle_{ab})\right],
$$
where the concrete expression of $\left|\psi_{i}\right\rangle$ is presented in (\ref{A}).
Then Alice can perform a three-outcome measurement, each outcome corresponding to an operator as
$$
\begin{aligned}
	M_1=|0\rangle_a\langle0|\otimes|0\rangle_A\langle0|+r^2|1\rangle_a\langle1|\otimes(|1\rangle_A\langle1|+|2\rangle_A\langle 2|),\\
    M_2=r^2|1\rangle_a\langle1|\otimes|0\rangle_A\langle0|+|0\rangle_a\langle0|\otimes(|1\rangle_A\langle1|+|2\rangle_A\langle 2|),\\
    M_3=\sqrt{1-r^4}|1\rangle_a\langle1|\otimes I_A\;\;\;\;\;\;\;\;\;\;\;\;\;\;\;\;\;\;\;\;\;\;\;\;\;\;\;\;\;\;\;\;\;\;\;.
\end{aligned}
$$
satisfying the completeness equation $\sum_{i=1}^3 M_i^{\dagger} M_i=I_{a A}$, where $I_{a A}$ is the identity operator of the parts $a$ and $A$.  Alice will obtain the outcome $M_t$ with probability $\gamma_t=\left\langle\psi_{i}^{\prime}\left|\left(M_t^{\dagger} M_t\otimes I_{b B} \right)\right| \psi_{i}^{\prime}\right\rangle$, and the state of the system after the measurement is $\frac{\left(M_t\otimes I_{b B}\right)\left|\psi_{i k}^{\prime}\right\rangle}{\sqrt{\left\langle\psi_{i k}^{\prime}\left|\left(M_t^{\dagger} M_t\otimes I_{b B}\right)\right| \psi_{i k}^{\prime}\right\rangle}}$, where $\quad t=1,2,3$ and $I_{b B}$ is the identity operator of the parts $b$ and $B$.
The post measurement reduced states corresponding to the outcome $M_1$ is,
\begin{multline}
	\label{B}
	$$
	\left|\psi_{1},\psi_{2}\right\rangle\rightarrow|0\rangle_{A}|1\rangle_{B}\left|0 \pm 1\rangle_{C}|00\rangle_{ab}\right.,\\
	\left|\psi_{3},\psi_{4}\right\rangle\rightarrow
	|0\rangle_{A}|2\rangle_{B}\left|0 \pm 2\rangle_{C}|00\rangle_{ab}\right.,\;\;\;\;\;\;\;\;\;\;\;\;\;\;\;\;\;\;\;\;\;\;\;\;\;\;\;\;\;\;\\
	\left|\psi_{5},\psi_{6}\right\rangle\rightarrow|1\rangle_{A}\left|0 \pm
	1\rangle_{B}|0\rangle_{C}|11\rangle_{ab}\right.,\;\;\;\;\;\;\;\;\;\;\;\;\;\;\;\;\;\;\;\;\;\;\;\;\;\;\;\;\;\;\\
	\left|\psi_{7},\psi_{8}\right\rangle\rightarrow|2\rangle_{A}\left|0 \pm 2\rangle_{B}|0\rangle_{C}|11\rangle_{ab}\right.,\;\;\;\;\;\;\;\;\;\;\;\;\;\;\;\;\;\;\;\;\;\;\;\;\;\;\;\;\;\;\\
	\left|\psi_{9},\psi_{10}\right\rangle\rightarrow|0\rangle_{A}\left|0\rangle_{B}|1\rangle_{C}|00\rangle_{ab}\right.\pm|1\rangle_{A}\left|0\rangle_{B}|1\rangle_{C}|11\rangle_{ab}\right.,\\
	\left|\psi_{11},\psi_{12}\right\rangle\rightarrow|0\rangle_{A}\left|0\rangle_{B}|2\rangle_{C}|00\rangle_{ab}\right.\pm|2\rangle_{A}\left|0\rangle_{B}|2\rangle_{C}|11\rangle_{ab}\right.,\\
	$$
\end{multline}
and the outcome $M_1$ occur with probability $\mathcal{P}_1=\frac{r^4}{1+r^4}$. Afterthat, Bob performs a measurement
\begin{multline*}
$$
	\mathcal{B}\equiv\{N_1:=\mathbb{P}\left[|1\rangle_{B} ;|0\rangle_{b}\right],N_2:=\mathbb{P}\left[|2\rangle_{B} ;|0\rangle_{b}\right],\\\bar{N}:=\mathbb{I}-N_1-N_2\}
$$
\end{multline*}
If the outcomes corresponding to $N_1$ and $N_2$ clicks. The resulting post measurement states are therefore $\cos\theta|0\rangle_{A}|1\rangle_{B}\left|0 \pm 1\rangle_{C}|00\rangle_{ab}\right.$ and $\cos\theta|0\rangle_{A}|2\rangle_{B}\left|0 \pm 2\rangle_{C}|00\rangle_{ab}\right.$ respectively, which can be easily distinguished by Charlie by projecting onto $|0\pm1\rangle_{C}, |1\pm2\rangle_{C}$. If the outcome corresponding to $\bar{N}$ clicks, then the remaining eight states are isolated which can be easily distinguished by Charlie and Alice.\\
\begin{figure}[h!]
	\centering
	\includegraphics[width=0.37\textwidth]{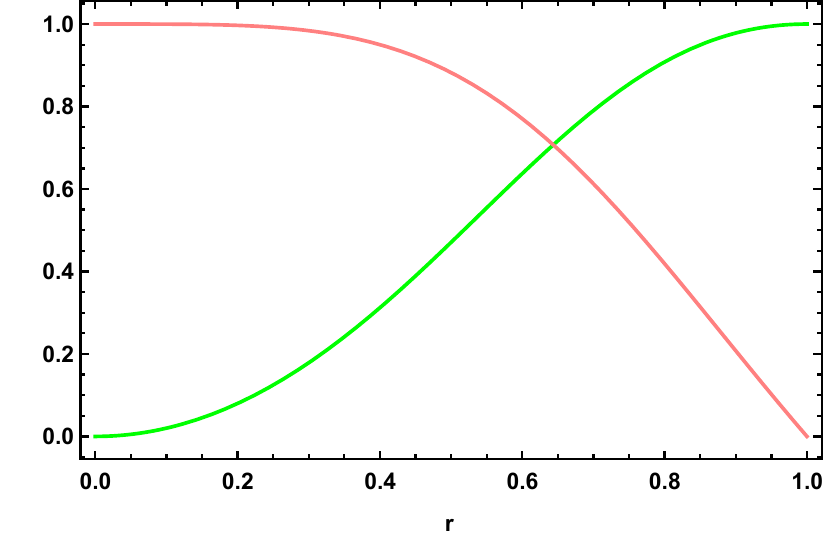}
	\caption{ A nonmaximally entangled state $|\chi\rangle=\frac{r}{\sqrt{1+r^4}}(r|00\rangle_{ab}$+$\frac{1}{r}|11\rangle_{ab})$, |r|$\leq$1, shared between Alice and Bob. The quantity of concurrence can be calculated by $\mathcal{C}_{AB}{(\chi)}=\frac{2r^2}{1+r^4}$(the green outline) and the probability of error can be calculated by $p_e=\frac{1-r^4}{1+r^4}\xi_e, 0<\xi_e<1$(the red outline).}
\label{f1}
\end{figure}
Alice obtains the second outcome $M_2$ with probability $\mathcal{P}_2=\frac{r^4}{1+r^4}$, then this creates new states which differ from the states (\ref{B}) only by ancillary systems $|00\rangle_{ab}\rightarrow|11\rangle_{ab}$ and
$|11\rangle_{ab}\rightarrow|00\rangle_{ab}$.\\
Alice obtains the third outcome $M_3$ with probability $\mathcal{P}_3=\frac{1-r^4}{1+r^4}$, and the post measurement states due to $M_3$ is, 
\begin{multline}
	$$
	\left|\psi_{1},\psi_{2}\right\rangle\rightarrow|0\rangle_{A}|1\rangle_{B}\left|0 \pm 1\rangle_{C}|11\rangle_{ab}\right.,\\
	\left|\psi_{3},\psi_{4}\right\rangle\rightarrow|0\rangle_{A}|2\rangle_{B}\left|0 \pm 2\rangle_{C}|11\rangle_{ab}\right.,\\
	\left|\psi_{5},\psi_{6}\right\rangle\rightarrow|1\rangle_{A}\left|0 \pm
	1\rangle_{B}|0\rangle_{C}|11\rangle_{ab}\right.,\\
	\left|\psi_{7},\psi_{8}\right\rangle\rightarrow|2\rangle_{A}\left|0 \pm 2\rangle_{B}|0\rangle_{C}|11\rangle_{ab}\right.,\\
	\left|\psi_{9},\psi_{10}\right\rangle\rightarrow|0\pm1\rangle_{A}\left|0\rangle_{B}|1\rangle_{C}|11\rangle_{ab}\right.,\\
	\left|\psi_{11},\psi_{12}\right\rangle\rightarrow|0\pm2\rangle_{A}\left|0\rangle_{B}|2\rangle_{C}|11\rangle_{ab}\right.,\\
	$$
\end{multline}
It is to be noted that the post-measurement reduced states have not been changed from (\ref{A}) by omitting the ancillary parts $|11\rangle_{ab}$ due to the outcome $M_3$ It is worthy of attention that with the measurement outcome $M_3$, the initial $12$ states have not been changed as (\ref{A}) by omitting the ancillary parts $|11\rangle_{a b}$. So, we can continue to do some measurements to distinguish them without ancillary entanglement system in probability, where an ambiguous discrimination strategy always can be taken, or we can simply guess which state is prepared in the system. Here we denote the probability of error by $\xi_e(0<\xi_e<1)$, and we do not need to known the concrete expression of $\xi_e$.

To distinguish these states, we first let Alice make a measurement. If Alice obtains the first outcome $M_1$ or the second outcome $M_2$, these states after the measurement can be perfectly distinguished by LOCC. If Alice obtains the third outcome $M_3$, we can continue to distinguish them without entanglement. Thus, the final probability of error is $p_e=\mathcal{P}_1\times0+\mathcal{P}_2\times 0+\mathcal{P}_3 \times \tau_e$=$\frac{1-r^4}{1+r^4}\xi_e, 0<\xi_e<1$. This completes the proof.$\blacksquare$ \\

We now explore the relationship between the probability of error in state discrimination and the quantity of entanglement, using a two-qubit non-maximally entangled state. Consider the state 
$|\chi\rangle=\frac{r}{\sqrt{1+r^4}}(r|00\rangle_{ab}$+$\frac{1}{r}|11\rangle_{ab})$, where $|r| \leq 1$. The degree of entanglement in this state can be quantified using concurrence, which is a widely used measure of entanglement for two-qubit systems. The concurrence $\mathcal{C}_{AB}(\chi)$ for the given state is given by the formula $\mathcal{C}_{AB}{(\chi)}=\frac{2r^2}{1+r^4}$.
This expression shows how the concurrence, and thus the entanglement, depends on the parameter $r$. When $r = 1$, the state becomes maximally entangled, and the concurrence reaches its maximum value of 1. As $r$ decreases, the concurrence decreases as well, indicating a reduction in entanglement. This relationship between $r$ and the concurrence will be key in analyzing how entanglement influences the probability of error in various quantum tasks, such as state discrimination.

According to Theorem~\ref{th1}, the probability of error in distinguishing the orthogonal states can be expressed as 
$p_e=\frac{1-r^4}{1+r^4}\xi_e, 0<\xi_e<1$, where $\xi_e$ is a parameter that modulates the error probability. The exact behavior of these functions is illustrated in Fig.~\ref{f1}. From the figure, it is evident that as the amount of entanglement shared between Alice and Bob decreases, the probability of error in locally distinguishing the orthogonal states described by Eq.~(\ref{A}) increases. Notably, when $r = 1$, the entanglement reaches its maximum value of one ebit, and the probability of error becomes zero. This demonstrates that perfect entanglement enables error-free discrimination, while reduced entanglement increases the likelihood of error in the discrimination process.

\textbf{Corollary 2} {\it If $|\chi\rangle=|\mathbb{\psi}(\frac{\pi}{2},\frac{\pi}{2},\frac{\pi}{2},r)\rangle$, the set of states (\ref{A}) can be distinguished with probability of error $p_e=\frac{1-r^4}{1+r^4}\xi_e, 0<\xi_e<1$.}\\\\
Throughout the paper we are dealing with a three outcome measurement set up denoted as $\mathcal{M}(\mathcal{X},\mathcal{Y})$ and defined by,
\begin{multline*}
$$
\mathcal{M}(\mathcal{X},\mathcal{Y}) \equiv\left\{M_1:=\mathcal{X}\mathbb{P}\left[|0\rangle_{A} ;|0\rangle_{a}\right]+\mathcal{Y}\mathbb{P}\left[(|1\rangle,|2\rangle)_{A} ;|1\rangle_{a}\right]\right.,\;\;\;\;\;\;\;\;\;\;\;\;\;\;\;\;\;\;\\
M_2:=\mathcal{Y}\mathbb{P}\left[|0\rangle_{A} ;|1\rangle_{a}\right]+\mathcal{X}\mathbb{P}\left[(|1\rangle,|2\rangle)_{A} ;|0\rangle_{a}\right],\\
M_3:=I_A\otimes( \sqrt{1-\mathcal{X}^2}\mathbb{P}[|0\rangle_{a}]+\sqrt{1-\mathcal{Y}^2}\mathbb{P}[|1\rangle_{a}])\}
$$
\end{multline*}
satisfying the completeness equation $\sum_{i=1}^3 M_i^{\dagger} M_i=I_{a A}$, where $I_{A}$ is the identity operator of the parts $A$.\\
\begin{figure}[h!]
	\centering
	\includegraphics[width=0.39\textwidth]{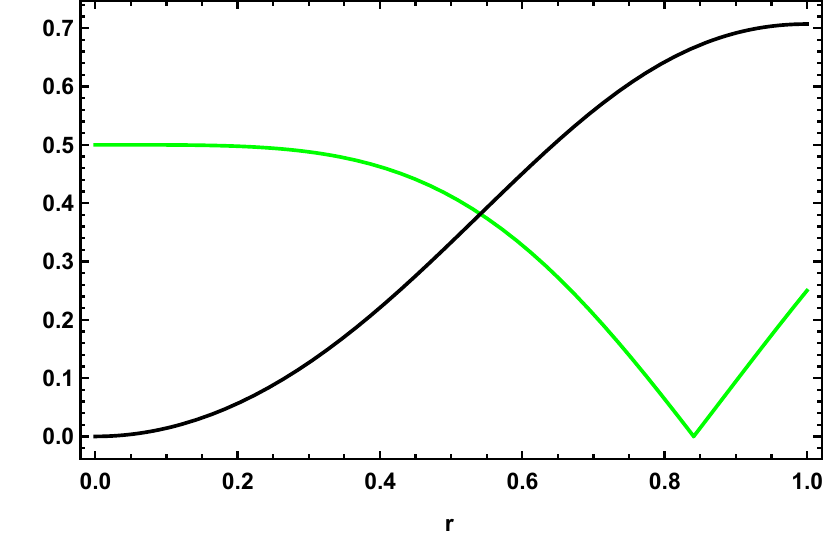}
	\caption{ A nonmaximally entangled state $|\chi\rangle=|\mathbb{\psi}(\frac{\pi}{2},\frac{\pi}{2},\frac{\pi}{4},r)\rangle$, |r|$\leq$1, shared between Alice, Bob and Charlie. The quantity of concurrence can be calculated by $\mathcal{C}_{AB}(|\chi\rangle)=\frac{\sqrt{2}|r|^2}{1+|r|^4}$(the black outline) and the probability of error can be calculated by $p_e$ (the green outline).}
	\label{f2}
\end{figure}
The non-generic GHZ class refers to a special category of GHZ states, which are maximally entangled quantum states involving multiple parties (typically three or more). In this non-generic class, the entanglement properties deviate from the typical GHZ structure, leading to more specific patterns of entanglement between subsystems. In particular, while the state may exhibit bipartite entanglement between certain pairs of subsystems, other pairs might show no entanglement at all. This creates an asymmetry in the distribution of entanglement, where the entanglement is not equally shared among all parties. The complexity of the system's quantum correlations gradually increases as parameters (like angles) evolve, reflecting more intricate quantum behavior.
\begin{theorem}
	\label{th2}
if $|
\chi\rangle=|\mathbb{\psi}(\frac{\pi}{2},\frac{\pi}{2},\frac{\pi}{4},r)\rangle$,|r|$\leq$1.\\
($\mathsf{i}$) the set of states (\ref{A}) can be distinguished with probability of error $$p_e=\begin{Bmatrix}
	\frac{1}{2(1+r^4)}(1-2r^4)\xi_e, & r^4\leq\frac{1}{2}\\
	\frac{r^4}{1+r^4}(1-\frac{1}{2r^4})\xi_e, & r^4>\frac{1}{2} 
\end{Bmatrix}$$ where 0<$\xi_e$<1.\\
($\mathsf{ii}$) $\mathcal{C}_{AB}(|\chi\rangle)=\frac{\sqrt{2}|r|^2}{1+|r|^4}$, $\mathcal{C}_{AC}(|\chi\rangle)$=$\mathcal{C}_{BC}(|\chi\rangle)$=0.\\
($\mathsf{iii}$) $\mathcal{T}_{ABC}(|\chi\rangle)=\frac{\sqrt{2}|r|^4}{(1+|r|^4)^2}$,\\
\end{theorem} 
{\it Proof.} Suppose an ancillary entangled state $|\mathbb{\psi}(\frac{\pi}{2},\frac{\pi}{2},\frac{\pi}{4},r)\rangle$ shared between Alice, Bob and Charlie. Then the initial states are transformed into
$$
\left|\psi_{i}^{\prime}\right\rangle=\left|\psi_{i}\right\rangle\otimes|\mathbb{\psi}(\frac{\pi}{2},\frac{\pi}{2},\frac{\pi}{4},r)\rangle,
$$
where the concrete expression of $\left|\psi_{i}\right\rangle$ is presented in (\ref{A}).
Then Alice performs a three-outcome measurement $$\mathcal{A}=\begin{Bmatrix}
	\mathcal{M}$$(1,\sqrt{2}r^2), & r^4\leq\frac{1}{2}\\
	\mathcal{M}$$(\frac{1}{\sqrt{2}r^2},1), & r^4>\frac{1}{2} 
\end{Bmatrix}$$ The post measurement reduced states corresponding to the outcome $M_1$ is,
\begin{multline*}
	$$
	\left|\psi_{1},\psi_{2}\right\rangle\rightarrow |0\rangle_{A}|1\rangle_{B}\left|0 \pm 1\rangle_{C}|000\rangle_{abc}\right.,\\
	\left|\psi_{3},\psi_{4}\right\rangle\rightarrow |0\rangle_{A}|2\rangle_{B}\left|0 \pm 2\rangle_{C}|000\rangle_{abc}\right.,\;\;\;\;\;\;\;\;\;\;\;\;\;\;\;\;\;\;\;\;\;\;\;\;\;\;\;\;\;\;\\
	\left|\psi_{5},\psi_{6}\right\rangle\rightarrow |1\rangle_{A}\left|0 \pm
	1\rangle_{B}|0\rangle_{C}|11\rangle_{ab}|0+1\rangle_{c}\right.,\;\;\;\;\;\;\;\;\;\;\;\;\;\;\;\;\;\;\;\;\;\;\;\;\;\;\;\;\;\;\\
	\left|\psi_{7},\psi_{8}\right\rangle\rightarrow |2\rangle_{A}\left|0 \pm2
	\rangle_{B}|0\rangle_{C}|11\rangle_{ab}|0+1\rangle_{c}\right.,\;\;\;\;\;\;\;\;\;\;\;\;\;\;\;\;\;\;\;\;\;\;\;\;\;\;\;\;\;\;\\
	\left|\psi_{9},\psi_{10}\right\rangle\rightarrow[|001\rangle_{ABC}|000\rangle_{abc}\pm|101\rangle_{ABC}|110\rangle_{abc}\\\pm|101\rangle_{ABC}|111\rangle_{abc}],\\
	\left|\psi_{11},\psi_{12}\right\rangle\rightarrow[|002\rangle_{ABC}|000\rangle_{abc}\pm|202\rangle_{ABC}|110\rangle_{abc}\\\pm|202\rangle_{ABC}|111\rangle_{abc}],\\
	$$
\end{multline*}
and the outcome $M_1$ occur with probability $$\mathcal{P}_1=\begin{Bmatrix}
	 \frac{r^4}{1+r^4}, & r^4\leq\frac{1}{2}\\
	\frac{1}{2(1+r^4)} & r^4>\frac{1}{2} 
\end{Bmatrix}$$ Afterthat, Bob and Charlie perform a sequence of measurement which is followed by previous theorem.\\
The impact of $M_2$ on (\ref{A}) gives,
\begin{multline*}
	$$
	\left|\psi_{1},\psi_{2}\right\rangle\rightarrow|0\rangle_{A}|1\rangle_{B}\left|0 \pm 1\rangle_{C}|11\rangle_{ab}|0+1\rangle_{c}\right.,\\
	\left|\psi_{3},\psi_{4}\right\rangle\rightarrow|0\rangle_{A}|2\rangle_{B}\left|0 \pm 2\rangle_{C}|11\rangle_{ab}|0+1\rangle_{c}\right.,\;\;\;\;\;\;\;\;\;\;\;\;\;\;\;\;\;\;\;\;\;\;\;\;\;\;\;\;\;\;\\
	\left|\psi_{5},\psi_{6}\right\rangle\rightarrow|1\rangle_{A}\left|0 \pm
	1\rangle_{B}|0\rangle_{C}|000\rangle_{abc}\right.,\;\;\;\;\;\;\;\;\;\;\;\;\;\;\;\;\;\;\;\;\;\;\;\;\;\;\;\;\;\;\\
	\left|\psi_{7},\psi_{8}\right\rangle\rightarrow|2\rangle_{A}\left|0 \pm2
	\rangle_{B}|0\rangle_{C}|000\rangle_{abc}\right.,\;\;\;\;\;\;\;\;\;\;\;\;\;\;\;\;\;\;\;\;\;\;\;\;\;\;\;\;\;\;\\
	\left|\psi_{9},\psi_{10}\right\rangle\rightarrow[|001\rangle_{ABC}|11\rangle_{ab}|0+1\rangle_{c}\pm|101\rangle_{ABC}|000\rangle_{abc}],\\
	\left|\psi_{11},\psi_{12}\right\rangle\rightarrow[|002\rangle_{ABC}|11\rangle_{ab}|0+1\rangle_{c}\pm|202\rangle_{ABC}|000\rangle_{abc}],\\
	$$
\end{multline*}
And it occur with probability$$\mathcal{P}_2=\begin{Bmatrix}
	 \frac{r^4}{1+r^4}, & r^4\leq\frac{1}{2}\\
\frac{1}{2(1+r^4)} & r^4>\frac{1}{2}  
\end{Bmatrix}$$also.
Next, Bob and Charlie perform a sequence of measurement which is followed by previous theorem.\\
Alice obtains the third outcome $M_3$ with probability $$p_e=\begin{Bmatrix}
	\frac{1}{2(1+r^4)}(1-2r^4), & r^4\leq\frac{1}{2}\\
\frac{r^4}{1+r^4}(1-\frac{1}{2r^4}), & r^4>\frac{1}{2} 
\end{Bmatrix}$$ and the post measurement states due to $M_3$ becomes, 
\begin{multline*}
	$$
	\left|\psi_{1},\psi_{2}\right\rangle\rightarrow|0\rangle_{A}|1\rangle_{B}\left|0 \pm 1\rangle_{C}|00\rangle_{ab}\right.,\\
	\left|\psi_{3},\psi_{4}\right\rangle\rightarrow|0\rangle_{A}|2\rangle_{B}\left|0 \pm 2\rangle_{C}|00\rangle_{ab}\right.,\\
	\left|\psi_{5},\psi_{6}\right\rangle\rightarrow|1\rangle_{A}\left|0 \pm
	1\rangle_{B}|0\rangle_{C}|00\rangle_{ab}\right.,\\
	\left|\psi_{7},\psi_{8}\right\rangle|2\rangle_{A}\left|0 \pm 2\rangle_{B}|0\rangle_{C}|00\rangle_{ab}\right.,\\
	\left|\psi_{9},\psi_{10}\right\rangle\rightarrow|0\pm1\rangle_{A}\left|0\rangle_{B}|1\rangle_{C}|00\rangle_{ab}\right.,\\
	\left|\psi_{11},\psi_{12}\right\rangle\rightarrow|0\pm2\rangle_{A}\left|0\rangle_{B}|2\rangle_{C}|00\rangle_{ab}\right.,\\
	$$
\end{multline*}
 Thus, the final probability of error is $p_e=\mathcal{P}_1\times0+\mathcal{P}_2\times 0+\mathcal{P}_3 \times \xi_e$, $\xi_e(0<\xi_e<1)$. This completes the proof.$\blacksquare$\\

We will now investigate how the probability of error correlates with the level of entanglement. With the three qubit nonmaximally entangled state $|
 \chi\rangle=|\mathbb{\psi}(\frac{\pi}{2},\frac{\pi}{2},\frac{\pi}{4},r)\rangle$, |r|$\leq$1. The quantity of concurrence can be calculated by the formula $\mathcal{C}_{AB}(|\chi\rangle)=\frac{\sqrt{2}|r|^2}{1+|r|^4}$, $\mathcal{C}_{AC}(|\chi\rangle)$=$\mathcal{C}_{BC}(|\chi\rangle)$=0.
 
 According to Theorem~\ref{th2}, we know that the probability of error can be calculated by $$p_e=\begin{Bmatrix}
 	\frac{1}{2(1+r^4)}(1-2r^4)\xi_e, & r^4\leq\frac{1}{2}\\
 	\frac{r^4}{1+r^4}(1-\frac{1}{2r^4})\xi_e, & r^4>\frac{1}{2} 
 \end{Bmatrix}$$ where 0<$\xi_e$<1. The exact values of these two functions are shown in Fig.~\ref{f2}. In the region $r\leq (\frac{1}{2})^{\frac{1}{4}}$ we can see that the less entanglement shared between Alice and Bob, the more the probability
 of error to locally distinguish the orthogonal set of states (\ref{A}). Moreover, with $r=(\frac{1}{2})^\frac{1}{4}$, the probability of error is zero; at the same time, the quantity of entanglement distribution is as follows, $\mathcal{C}_{AB}(|\chi\rangle)=\frac{2}{3}$, $\mathcal{C}_{AC}(|\chi\rangle)$=$\mathcal{C}_{BC}(|\chi\rangle)$=0.\\\\
\begin{figure}[h!]
	\centering
	\includegraphics[width=0.39\textwidth]{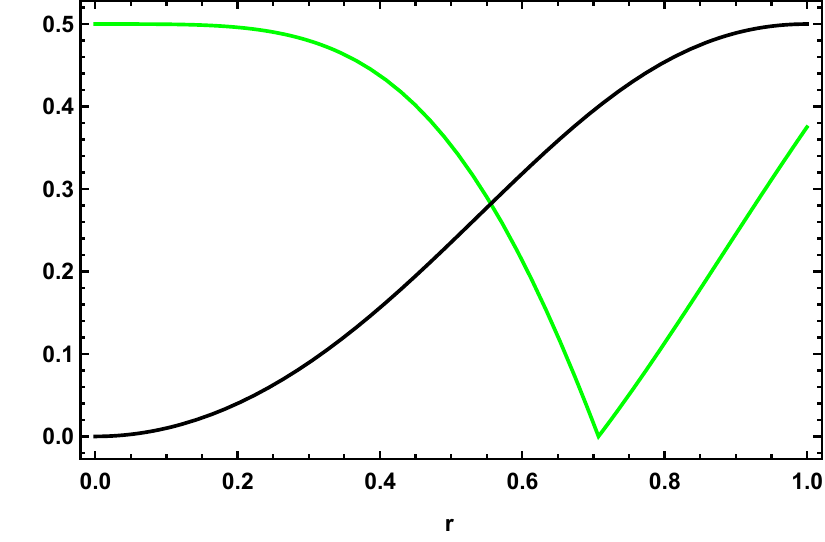}
	\caption{ A nonmaximally entangled state $|\chi\rangle=|\mathbb{\psi}(\frac{\pi}{2},\frac{\pi}{4},\frac{\pi}{4},r)\rangle$, |r|$\leq$1, shared between Alice, Bob and Charlie. The quantity of concurrence can be calculated by $\mathcal{C}_{AB}(|\chi\rangle)=\frac{|r|^2}{1+|r|^4}$(the black outline) and the probability of error can be calculated by $p_e$ (the green outline).}
\label{f3}
\end{figure}
\begin{theorem}
	\label{th3}
	If $|\chi\rangle=|\mathbb{\psi}(\frac{\pi}{2},\frac{\pi}{4},\frac{\pi}{4},r)\rangle$,\\
	($\mathsf{i}$) the set of states (\ref{A}) can be distinguished with probability of error $$p_e=\begin{Bmatrix}
		\frac{1}{4(1+r^4)}(1-4r^4)\xi_e, & r^2\leq\frac{1}{2} \\
		\frac{r^4}{1+r^4}(1-\frac{1}{4r^4})\xi_e, & r^2>\frac{1}{2}
	\end{Bmatrix}$$ where 0<$\xi_e$<1.\\
	($\mathsf{ii}$) $\mathcal{C}_{AB}(|\chi\rangle)=\frac{|r|^2}{1+|r|^4}$, $\mathcal{C}_{AC}(|\chi\rangle)=\frac{|r|^2}{1+|r|^4}$,$\mathcal{C}_{BC}(|\chi\rangle)$=0,\\
	($\mathsf{iii}$) $\mathcal{T}_{ABC}(|\chi\rangle)=\frac{|r|^4}{(1+|r|^4)^2}$.
\end{theorem}
{\it Proof.} Without lose of generality suppose an ancillary entangled state $|\mathbb{\psi}(\frac{\pi}{2},\frac{\pi}{4},\frac{\pi}{4},r)\rangle$ shared between Alice Bob and Charlie. Then the initial states are transformed into
$$
\left|\psi_{i}^{\prime}\right\rangle=\left|\psi_{i}\right\rangle\otimes|\mathbb{\psi}(\frac{\pi}{2},\frac{\pi}{4},\frac{\pi}{4},r)\rangle,
$$
where the concrete expression of $\left|\psi_{i}\right\rangle$ is presented in (\ref{A}). Alice can perform a three-outcome measurement,
$$\mathcal{A}=\begin{Bmatrix}
	\mathcal{M}$$(1,2r^2), & r^2\leq\frac{1}{2}\\
	\mathcal{M}$$(\frac{1}{2r^2},1), & r^2>\frac{1}{2} 
\end{Bmatrix}$$
satisfying the completeness equation $\sum_{i=1}^3 M_i^{\dagger} M_i=I_{a A}$, where $I_{a A}$ is the identity operator of the parts $a$ and $A$. Afterthat Bob and charlie do some sequence of measurements as described before. Thus, the final probability of error is $p_e=\mathcal{P}_1\times0+\mathcal{P}_2\times 0+\mathcal{P}_3 \times \xi_e$. This completes the proof.$\blacksquare$\\

 Now, we analyze the relationship between the probability of error and the quantity of entanglement. With the three qubit nonmaximally entangled state $|
\chi\rangle=|\mathbb{\psi}(\frac{\pi}{2},\frac{\pi}{4},\frac{\pi}{4},r)\rangle$, |r|$\leq$1. the quantity of concurrence can be calculated by the formula $\mathcal{C}_{AB}(|\chi\rangle)=\frac{|r|^2}{1+|r|^4}$, $\mathcal{C}_{AC}(|\chi\rangle)=\frac{|r|^2}{1+|r|^4}$, $\mathcal{C}_{BC}(|\chi\rangle)$=0.

According to Theorem~\ref{th3}, we know that the probability of error can be calculated by $$p_e=\begin{Bmatrix}
	\frac{1}{4(1+r^4)}(1-4r^4)\xi_e, & r^2\leq\frac{1}{2} \\
	\frac{r^4}{1+r^4}(1-\frac{1}{4r^4})\xi_e, & r^2>\frac{1}{2}
\end{Bmatrix}$$ where 0<$\xi_e$<1, where the exact values of these two functions are shown in Fig.~\ref{f3}. In the region $r\leq (\frac{1}{2})^{\frac{1}{2}}$, we can see that the less entanglement shared between Alice and Bob, the more the probability of error to locally distinguish the orthogonal set of states (\ref{A}). Moreover, with $r=(\frac{1}{2})^\frac{1}{2}$, the probability of error is zero; at the same time, the quantity of entanglement distribution is as follows, $\mathcal{C}_{AB}(|\chi\rangle)=\frac{2}{5}$=$\mathcal{C}_{AC}(|\chi\rangle)$, $\mathcal{C}_{BC}(|\chi\rangle)$=0.\\\\
\begin{figure}
	\includegraphics[width=0.39\textwidth]{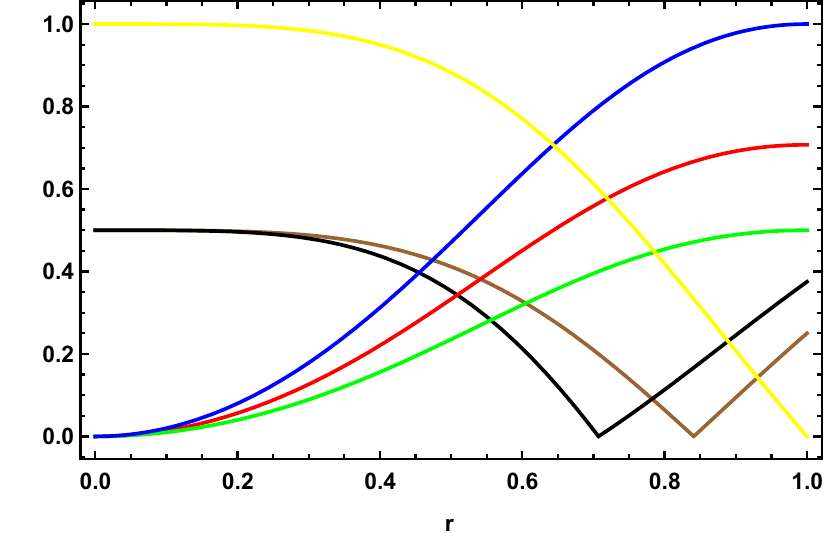}
	\caption{(i) The blue and yellow outlines represent the consumption of resource between (Alice,Bob) and the probability of error respectively when, $|\chi\rangle=\frac{r}{\sqrt{1+r^4}}(r|00\rangle_{ab}$+$\frac{1}{r}|11\rangle_{ab})$, |r|$\leq$1. (ii) The red and brown outlines represent the consumption of resource between (Alice,Bob) and the probability of error respectively when,$|\chi\rangle=|\mathbb{\psi}(\frac{\pi}{2},\frac{\pi}{2},\frac{\pi}{4},r)\rangle$,|r|$\leq$1. (iii) The green and black outlines represent the consumption of resource between (Alice,Bob) and the probability of error respectively when,$|\chi\rangle=|\mathbb{\psi}(\frac{\pi}{2},\frac{\pi}{2},\frac{\pi}{4},r)\rangle$,|r|$\leq$1.}
	\label{f4}
\end{figure}
In \cite{Halder2018} Halder et al, showed that the set of states (\ref{A}) is initially locally indistinguishable and to distinguish it perfectly one copy of Bell state shared between any two parties is sufficient as a resource. By Lemma 2, it is clear that one copy of three qubit GHZ state is also sufficient to distinguish perfectly the set of states (\ref{A}). Therefore for such discrimination task  we are not able to differentiate between the advantages of multiparty correlation and bipartite correlation when both of them are at maximal level. But dilution of both resource states make the gap of advantages visible.
\section{The Impact of the Generic Class on Probabilistic Distinguishability in Quantum Systems}
\label{A4}

In the study of tripartite entangled states, it is crucial to examine the correlations and entanglement structures within the subsystems. A particular class of non-generic GHZ-like states, such as the resource state \( |\psi(\frac{\pi}{2}, \frac{\pi}{4}, \frac{\pi}{4}, r)\rangle \), serves as a prime example of a tripartite entangled state exhibiting a less correlated structure. In this case, the entanglement distribution is such that there is bipartite entanglement between subsystems A-B and A-C, while no bipartite entanglement exists between subsystem B-C. This configuration aligns with what is referred to as case-III in entanglement classification, where specific pairs of subsystems exhibit entanglement while others do not. On the other hand, the resource state \( |\psi(\frac{\pi}{2}, \frac{\pi}{2}, \frac{\pi}{4}, r)\rangle \) represents another variation of the tripartite entangled state, where bipartite entanglement is only present between subsystems A-B, with no entanglement detected between either subsystem pair B-C or A-C. This configuration places the state in case-II, where entanglement is restricted to a single pair of subsystems. These examples highlight the rich diversity in the structure of tripartite entanglement and the importance of understanding the specific ways in which entanglement manifests in such systems, particularly in terms of reduced subsystems and their interdependencies. Such investigations are key to advancing our knowledge of the resources required for quantum information tasks and the potential for less correlated entangled states in quantum communication protocols.\\

For the case of $|
\chi\rangle=|\mathbb{\psi}(\frac{\pi}{2},\frac{\pi}{2},\frac{\pi}{4},r)\rangle$, |r|$\leq$1, the quantity of concurrence between A-B is given by $\mathcal{C}_{AB}(|\chi\rangle)=\frac{\sqrt{2}|r|^2}{1+|r|^4}$, and the probability of error can be calculated by $$p_e=\begin{Bmatrix}
	\frac{1}{2(1+r^4)}(1-2r^4)\xi_e, & r^4\leq\frac{1}{2} \\
	\frac{r^4}{1+r^4}(1-\frac{1}{2r^4})\xi_e, & r^4>\frac{1}{2}
\end{Bmatrix}$$ where 0<$\xi_e$<1.\\
Now representing the probability of error $p_e$ in terms of concurrence of the reduced system A-B ($C_{AB}$) and tangle ($\tau$) we get the following equations respectively.
\begin{equation}
	p_e=\begin{Bmatrix}
		\frac{1-\frac{\left(1-\sqrt{1-2{\mathcal{C}_{AB}}}\right)^2}{{\mathcal{C}_{AB}}^2}}{2\left\lbrace1+\left(1+\frac{1-\sqrt{1-2{\mathcal{C}_{AB}}^2}}{\sqrt{2}{\mathcal{C}_{AB}}}\right)^2 \right\rbrace }\xi_e, & \mathcal{C}_{AB}\leq\frac{2}{3} \\
		\frac{1-\frac{{\mathcal{C}_{AB}}^2}{\left(1-\sqrt{1-2{\mathcal{C}_{AB}}}\right)^2}}{1+\frac{2{\mathcal{C}_{AB}}^2}{\left(1-\sqrt{1-2{\mathcal{C}_{AB}}}\right)^2}}\xi_e, & \mathcal{C}_{AB}>\frac{2}{3}
	\end{Bmatrix}
\end{equation} \\

 \begin{equation}
 	p_e=\begin{Bmatrix}
 		\frac{\tau}{2(1-\sqrt{1-2\tau})}(1-\frac{2-2\tau-2\sqrt{1-2\tau}}{\tau})\xi_e, & \tau\leq\frac{4}{9} \\
 		\frac{1-\tau-\sqrt{1-2\tau}}{1-\sqrt{1-2\tau}}(1-\frac{\tau}{2-2\tau-2\sqrt{1-2\tau}})\xi_e, & \tau>\frac{4}{9}
 	\end{Bmatrix}
 \end{equation}
where 0<$\xi_e$<1.\\
\begin{figure}[h]
	\centering
	\includegraphics[width=0.39\textwidth]{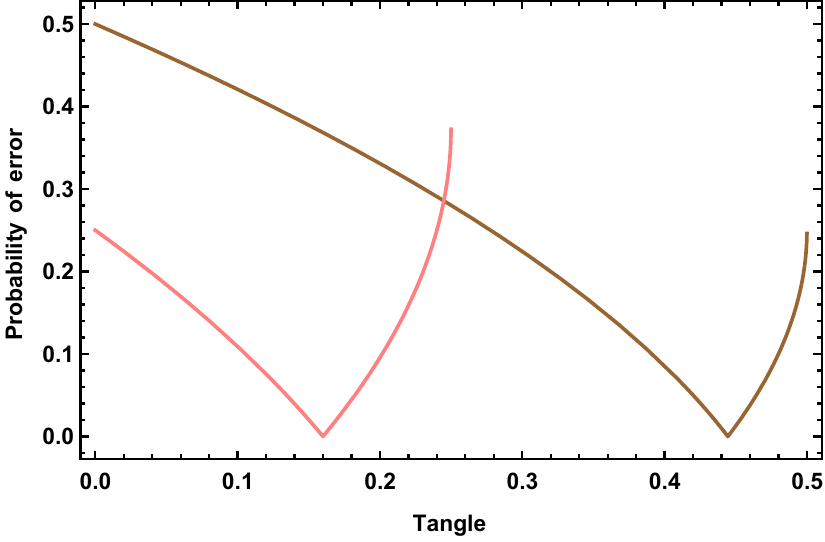}
	\caption{ This figure represents graph of $p_e$ with respect to $\tau$. Brown and pink outlines represent $p_e$ vs $\tau$ for $|\mathbb{\psi}(\frac{\pi}{2}, \frac{\pi}{2},\frac{\pi}{4},r)\rangle$ and $|\mathbb{\psi}(\frac{\pi}{2}, \frac{\pi}{4},\frac{\pi}{4},r)\rangle$, |r|$\leq$1 respectively.}
	\label{f6}
\end{figure}

\begin{figure}[h]
	\centering
	\includegraphics[width=0.39\textwidth]{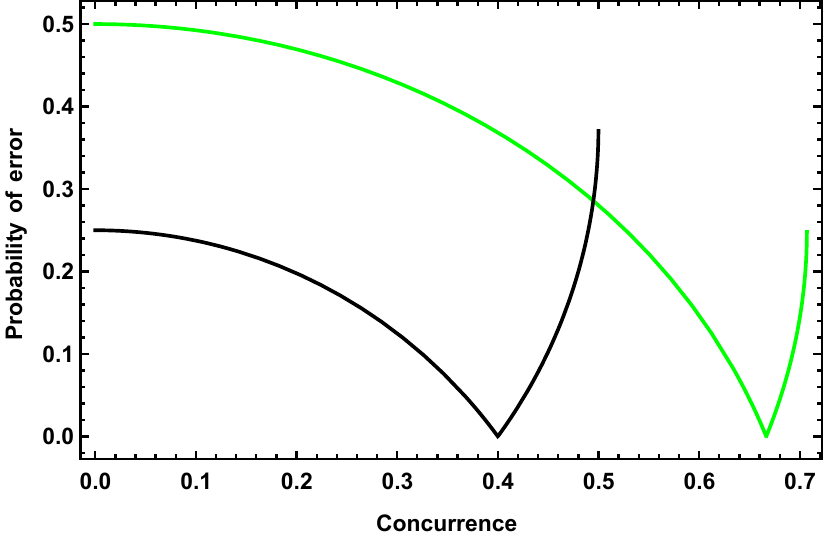}
	\caption{ This figure represents graph of $p_e$ with respect to $\mathcal{C}_{AB}$. Green and black outlines represent $p_e$ vs $\mathcal{C}_{AB}$ for $|\mathbb{\psi}(\frac{\pi}{2},\frac{\pi}{2},\frac{\pi}{4},r)\rangle$ and $|\mathbb{\psi}(\frac{\pi}{2},\frac{\pi}{4},\frac{\pi}{4}, r)\rangle$, |r|$\leq$1 respectively.}
	\label{f5}
\end{figure}

For the case of $|\chi\rangle=|\mathbb{\psi}(\frac{\pi}{2},\frac{\pi}{4},\frac{\pi}{4},r)\rangle$, |r|$\leq$1, the quantity of concurrence between A-B is given by $\mathcal{C}_{AB}(|\chi\rangle)=\frac{|r|^2}{1+|r|^4}$, and the probability of error can be calculated by $$p_e=\begin{Bmatrix}
	\frac{1}{4(1+r^4)}(1-4r^4)\xi_e, & r^2\leq\frac{1}{2} \\
	\frac{r^4}{1+r^4}(1-\frac{1}{4r^4})\xi_e, & r^2>\frac{1}{2}
\end{Bmatrix}$$ where 0<$\xi_e$<1.\\
Again representing the probability of error $p_e$ in terms of $C_{AB}$ and $\tau$ we get the following equations. 
\begin{equation}
	p_e=\begin{Bmatrix}
		\frac{1-\frac{\left(1-\sqrt{1-4{\mathcal{C}_{AB}}}\right)^2}{{\mathcal{C}_{AB}}^2}}{4\left\lbrace1+\left(1+\frac{1-\sqrt{1-4{\mathcal{C}_{AB}}^2}}{2{\mathcal{C}_{AB}}}\right)^2 \right\rbrace }\xi_e, & \mathcal{C}_{AB}\leq\frac{2}{5} \\
		\frac{1-\frac{{\mathcal{C}_{AB}}^2}{\left(1-\sqrt{1-4{\mathcal{C}_{AB}}}\right)^2}}{1+\frac{4{\mathcal{C}_{AB}}^2}{\left(1-\sqrt{1-4{\mathcal{C}_{AB}}}\right)^2}}\xi_e, & \mathcal{C}_{AB}>\frac{2}{5}
	\end{Bmatrix}
\end{equation} \\

\begin{equation}
	p_e=\begin{Bmatrix}
		\frac{\tau}{2(1-2\sqrt{1-4\tau})}(1-\frac{2-4\tau-4\sqrt{1-4\tau}}{\tau})\xi_e, & \tau\leq\frac{4}{25} \\
		\frac{1-2\tau-2\sqrt{1-4\tau}}{1-2\sqrt{1-4\tau}}(1-\frac{\tau}{2-4\tau-4\sqrt{1-4\tau}})\xi_e, & \tau>\frac{4}{25}
	\end{Bmatrix}
\end{equation}
where 0<$\xi_e$<1.\\
From Fig.\ref{f6} and Fig.\ref{f5}, we observe that the state \( |\mathbb{\psi}(\frac{\pi}{2}, \frac{\pi}{4}, \frac{\pi}{4}, r)\rangle \) offers a greater advantage for the distinguishing task outlined in equation (\ref{A}) compared to the state \( |\mathbb{\psi}(\frac{\pi}{2}, \frac{\pi}{2}, \frac{\pi}{4}, r)\rangle \). Specifically, this advantage becomes particularly evident when the values of \( \mathcal{C}_{AB} \) are less than or equal to approximately 0.495, and the parameter \( \tau \) is less than or equal to approximately 0.245. This suggests that the first state exhibits superior performance in terms of probabilistic distinguishability under these conditions, making it a more effective resource for such tasks in the specified parameter ranges. The visual representation in the figures helps us to clearly delineate these regions where one state outperforms the other in terms of its ability to distinguish between different quantum states.\\\\
Also from Fig.\ref{f6} and Fig.\ref{f5} it can be seen that for type-II state the probability of error is zero if the value of concurrence and tangle equals $\frac{2}{3}$ and $\frac{4}{9}$ respectively. Whereas, for the case of type-III the probability of error becomes zero when the value of concurrence and tangle equals $\frac{2}{5}$ and $\frac{4}{25}$ respectively. Now it is well known that type-III state is lesser entangled than type-II, as using LOCC we can always reach \cite{tarun}. So it is evident that a less entangled state provides us same result than a more entangled state. It is also interesting that for case of type-I state the probability of error is zero only when it is a GHZ state, which is the maximally entangled state in GHZ-SLOCC class. We conjecture that a lesser entangled state can provide error free distinguishability only due to its distribution of entanglement. Type-III state has two non-zero bipartite entangled reduced subsystems, whereas type-II and GHZ state has one and zero reduced subsystems with bipartite entanglement.\\
In \cite{Halder2018} authors showed that to distinguish (\ref{A}) perfectly one-copy of Bell state is sufficient.There is no such result till now that a non-maximally bipartite entangled state can be used for perfect distinguishability. In this paper we showed that there are three different genuine entangled states in GHZ- SLOCC class which we can use as a resource in the perfect discrimination of (\ref{A}). One of which is GHZ state, the maximally entangled state of GHZ-SLOCC class but the rest two are non-maximally entangled. Thus we can conclude that a non-maximally genuine multiparty entangled state become more powerful than a non-maximal bipartite entangled state for this discrimination task.   
 \section{discussion}
\label{A5}
	Multipartite entanglement is a key concept in quantum physics that captures the complex correlations between multiple quantum particles. It has practical uses in quantum technologies and is crucial for gaining insight into the behavior of complex quantum systems. In this paper, we address a notable gap in the current understanding of quantum state discrimination by investigating the discriminative potential of non-maximally entangled states within the three-qubit GHZ SLOCC class. Unlike existing research, which mainly relies on maximally entangled states for perfect discrimination, our proposed method explores the use of non-maximally entangled states as resources. Imposing the structural characteristics of the GHZ SLOCC class, we develop a systematic approach to examine the local distinguishability of orthogonal product states. Our methodology involves a detailed analysis of the relationship between the probability of error in discrimination and the quantity of entanglement present in bipartite and multipartite subsystems. By considering entanglement configurations in terms of bipartite cuts and multiparty entanglement, we aim to shed light on the interplay between entanglement classification and the local distinguishability of orthogonal product states. The outcomes of our investigation promise to enhance our understanding of quantum information processing and the intricate connection between entanglement and nonlocality in the realm of quantum mechanics. In our investigation, we reveal a significant advantage in the probabilistic distinguishability of orthogonal product states by employing the generic class of GHZ states. The unique properties of GHZ states within this class serve as a versatile and powerful resource, enhancing the ability to distinguish between states in probabilistic settings. This highlights the value of GHZ states in improving the discrimination process in complex quantum systems. Our findings further demonstrate that the generic class of GHZ states provides a distinct advantage in tackling the challenges related to the probabilistic distinguishability of orthogonal product states. The inherent qualities of these states make them particularly effective in overcoming the limitations typically encountered in distinguishing such states, offering a valuable resource for improving quantum discrimination techniques. 
\section*{ACKNOWLEDGEMENTS}
The authors I. Chattopadhyay and D. Sarkar acknowledge DST-FIST India. The author A. Bhunia, acknowledges Dr. R. Sengupta, IISER, Berhampur, India for valuable comments.

\end{document}